\documentclass[prd,aps,superscriptaddress,nofootinbib,amsmath,amssymb,showpacs]{revtex4}
\usepackage{graphicx}
\usepackage{dcolumn}
\usepackage{bm}
\usepackage{multirow}
\usepackage{adjustbox}
\usepackage{textcomp}
\usepackage{mathtext}
\usepackage{hyperref}
\graphicspath{{./eps/}}

\begin{document}
\title{\Large Editorial}
 \title{{\Large Neutrinos and their interactions}}

\author{M. Sajjad \surname{Athar}}
\email{sajathar@gmail.com}
\affiliation{Department of Physics, Aligarh Muslim University, Aligarh-202002, India}
\author{S. K. \surname{Singh}}
\affiliation{Department of Physics, Aligarh Muslim University, Aligarh-202002, India}

\begin{abstract}
We present a short overview of the importance of the study of "Neutrino Interactions in the Intermediate and High Energy Region", with an introduction to the neutrinos and a very brief description about the collection of invited articles.
\end{abstract}
%
\maketitle

\section*{}
\label{intro}
The story of neutrino physics, which started in 1930 with the hypothesis of Pauli’s “neutron”~\cite{Pauli}, assumed to be a massless, 
chargeless fermion of spin $\frac{1}{2}$, in order to explain the two outstanding problems of that time associated with the conservation laws {\it viz.} the conservation of energy and the conservation of angular momentum, has been an amazing one. This hypothesis got a solid foundation in 1933 with the theory of beta decay~\cite{Fermi} 
propounded by Fermi who rechristened the Pauli’s “neutron” to "neutrino" and argued that the four point interaction vertex in a beta decay is vectorial in nature. 
 Later with the observations of parity violation in beta decay~\cite{Wu}, and the observation of neutrinos to be left handed particles~\cite{Goldhaber},
  it was established that the weak interaction vertex has V-A(Vector-Axial Vector) nature, and a theory of neutrino interaction with matter was formulated using chiral($\gamma_5$) invariance assuming neutrinos to be massless~\cite{Feynman:1958ty}. To avoid ultraviolet catastrophe in $\nu_e - e$ scattering, it was assumed that the interaction is mediated by a heavy boson. In 1956, 
  electron type neutrino~(rather an antineutrino $\bar\nu_e$) was detected at the Savannah River reactor~\cite{Reines}. Later it was observed that there are three different flavors of neutrinos {\it viz.} electron neutrino~($\nu_e$), muon neutrino~($\nu_\mu$) and tauon neutrino~($\nu_\tau$), which are characterized by their own lepton quantum number $L_e$, $L_\mu$ and $L_\tau$, and these are conserved
   separately in the weak interaction. In the Standard Model~(SM) of particle physics~\cite{Glashow:1961tr,Weinberg:1967,Salam} presently known to best describe the properties of the fundamental particles and their interactions, there  are three generations of lepton flavors, each one of them placed in a weak isospin doublet, where corresponding to each charged lepton i.e. $e^-$, $\mu^-$ and 
   $\tau^-$, there is a massless neutrino of the same lepton number i.e. $\nu_e$, $\nu_\mu$ and $\nu_\tau$. 
   The existence of three flavors of neutrinos was experimentally established in 1989 when the Large Electron-Positron Collider(LEP) confirmed the presence of three active neutrinos~\cite{Mele:2015etc}. The interactions of these neutrinos with matter are described by the standard model gauge bosons {\it viz.} $W^+$,  $W^-$, and $Z^o$. The absolute masses of neutrinos are not known, and there are upper experimental limits on them obtained from the end point spectrum of beta decay for $\nu_e$, pion decay for $\nu_\mu$ and tauon decay for $\nu_\tau$, as well as from some other indirect methods.

The history of the progress of our understanding of the physics of neutrinos is full of surprises as they continue to challenge our expectations regarding the validity of certain symmetry principles and conservation laws in particle physics. Today, we know that neutrinos are the most abundant particles in the Universe after photons, but least understood, due to their weakly interacting nature, though they play important role not only in particle and nuclear physics, but also in cosmology and astrophysics. There are natural  sources of neutrinos like the ones produced during the nuclear fusion inside the star’s core, supernova bursts, decay of secondary cosmic ray particles in the earth’s atmosphere, geoneutrinos produced in the earth's core, etc., as well as the artificial sources of neutrinos like those produced from the nuclear reactors and the particle accelerators. Many of these neutrino sources are being used in learning the properties of neutrinos and their interactions with matter. These neutrinos are also helpful in knowing the various astrophysical information, about the sun’s core, composition of earth’s core, time and place of supernova explosion, etc.~\cite{S}.

The observations of solar neutrino anomaly and the atmospheric neutrino puzzle, are generally understood on the basis of neutrino flavor oscillation, a quantum mechanical effect which implies that at least two of these neutrinos have tiny masses. The observation of the  phenomenon of neutrino oscillation therefore requires new physics Beyond the Standard Model~(BSM). Neutrino oscillation has also been observed in the accelerator as well as reactor neutrino experiments. The three neutrino flavor states $\nu_e$, $\nu_\mu$ and $\nu_\tau$ of the standard model 
are considered to be mixture of the three mass eigenstates $m_1$, $m_2$ and $m_3$. The mixing is described in terms of the 
 Pontecorvo, Maki, Nagakawa, Sakata~(PMNS) matrix~\cite{Maki}, which is most popularly parameterized in terms of the
  three mixing angles $\theta_{12}$, $\theta_{13}$ and $\theta_{23}$, and a phase $\delta$, better known as $\delta_{CP}$ as it can be used to describe CP violation. Some of these oscillation parameters have been determined in the solar, reactor~($\theta_{12}$), accelerator~($\theta_{13}$) and atmospheric~($\theta_{23}$) neutrino experiments. One important determination which is yet to be made is whether the neutrinos follow normal mass hierarchy~($m_1~<~m_2~<~m_3$) or inverted mass hierarchy~($m_3~<~m_1~<~m_2$). This is because the neutrino oscillation experiments can determine only the square of the mass differences ${\Delta m}^2_{21}$~(sensitive solar and reactor sources) and the absolute value of $|{\Delta m}^2_{31}|$~(sensitive to reactor, accelerator and atmospheric), and the sign of ${\Delta m}^2_{31}$ is required to settle the
mass hierarchy problem. Recently, there is some information on $\delta_{CP}$, but it is very limited. To understand the properties of neutrinos as well as to determine the various parameters of the PMNS matrix, CP violating phase 
$\delta_{CP}$, and to determine the mass hierarchy in neutrino mass eigen states, several experiments in the low energies~(corresponding to reactor, solar, and supernova neutrinos) as well as in the medium energies~(corresponding to accelerator and atmospheric neutrinos) are being performed. 

In the region of very low energy relevant for the reactor and solar neutrinos, the exclusive transitions to the ground state or a few low excited states in the final nucleus are accessible. In the medium and high energy region, the accelerator experiments like MiniBooNE, T2K, SciBooNE, CNGS, OPERA, NOvA, etc., as well as atmospheric neutrino experiments like SuperKamiokande, IceCube, etc.,  have (anti)neutrinos in the energy range which is sufficient enough to excite many nuclear states, the energies are sufficient enough to create new particles and can induce various inclusive processes of quasielastic(QE), inelastic(IE) (like 1$\pi$, 1$\eta$, 1$K$, $YK$($Y=\Lambda, \Sigma$) production, etc.) and deep inelastic(DIS) scattering processes given by~\cite{S}:
 \begin{eqnarray}\label{Ch-4:process1_nu}
 \nu_{l} (\bar{\nu}_{l})   + N &\longrightarrow & l^{-} (l^{+})  + N^{\prime}, \quad \quad \rm{(QE)}\nonumber\\
 \bar{\nu}_{l} + N &\longrightarrow & l^{+} + Y,\;\; \rm{QE \;\;hyperon(Y) \;\;production}\nonumber\\
 \nu_{l} (\bar{\nu}_{l}) + N &\longrightarrow & l^{-} (l^{+}) + N^{\prime} + X,\;\;(IE), \;\;X=\pi, K, \eta,\nonumber\\
 \nu_{l} (\bar{\nu}_{l}) + N &\longrightarrow & l^{-} (l^{+}) + Y + K(\bar K),\quad \quad (IE \;\; \rm{Associated\;\; particle \;\;production})\nonumber\\
 \nu_{l} (\bar{\nu}_{l}) + N&\longrightarrow & l^{-} (l^{+}) + \rm{jet\;\; of\;\; hadrons\;\; (DIS)}, \rm{where}\nonumber\\
 &&N,N^\prime=n \;\;or\;\; p. \;\;\nonumber
\end{eqnarray}
 
\begin{figure}
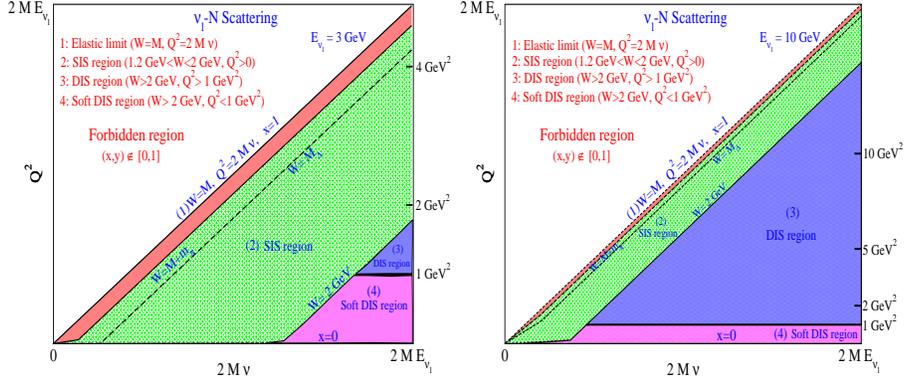

\begin{center}
\includegraphics[height=5 cm, width=5.9cm]{3gev_color_presentation.eps}
\includegraphics[height=5 cm, width=5.9cm]{10gev_color_presentation.eps}
 \end{center}
\caption{Allowed kinematical region for $\nu_l~-~N$ scattering in the ($Q^2, \nu$) plane for $E_\nu$=3 GeV(left panel) and 
$E_\nu$=10 GeV(right panel) for $Q^2 \ge 0$ ($Q^2$ is the four momentum transfer square). Invariant mass square is defined as $W^2=M^2+2M\nu-Q^2$ and the elastic limit is 
$x=\frac{Q^2}{2M_N\nu}=1$. The forbidden region in terms of $x$ and $y=\frac{\nu}{E_\nu}=\frac{(E_\nu - E_l)}{E_\nu}$ is defined 
as $x,y~\notin~[0,1]$. The process like photon emission is possible in the extreme left band(the region between $W=M$ and $W < M+m_\pi$). The SIS region has been defined as the region for which $M+m_\pi \le W \le 2GeV$ and $Q^2 \ge 0$, the DIS 
region is defined as the region for which $Q^2 \ge 1~GeV^2$ and $W \ge 2~GeV$, and Soft DIS region is defined as $Q^2 < 1GeV^2$ 
and $W \ge 2~GeV$. The soft DIS region is also nothing but the SIS region. The extreme left band also gets the contribution for the bound nucleons in nuclear targets through $np-nh$ like 2p-2h 
excitations. The boundaries between regions are not sharply established and are indicative only.} \label{fig3}
\end{figure}
 
  This volume is focused on the interaction of the intermediate and high energy neutrinos in the region of a few GeV. The inclusive cross sections in the quasielastic region are analysed in terms of the weak form factors of the nucleon and the cross sections in the 
  inelastic scattering corresponding to the excitations of various
nucleon resonances, lying in the first or higher
resonance region are described in terms of the transition form factors corresponding to the nucleon to resonance transition. On the other hand, if the energy transfer($\nu$) and the four momentum transfer square($Q^2$) are large, the inclusive cross sections
are expressed in terms of the structure functions corresponding to the deep inelastic scattering
process from the quarks and gluons in the nucleon. In the intermediate energy region corresponding to the
transition between resonance excitations and DIS, we are yet to find a method best suited
to describe the inclusive charged lepton or (anti)neutrino scattering processes. Using the
kinematical cuts in the $Q^2 - \nu$ plane (Figure \ref{fig3}), one may understand the regions like elastic scattering
($W = M$), inelastic scattering($M \le W \le 2$~GeV), deep inelastic scattering($Q^2 > 1 GeV^2$, $W >$ 2 GeV) and the soft DIS ($Q^2 < 1
GeV^2$, $W > 2 GeV$) regions. 

In the quasielastic region, for the scattering of (anti)neutrinos with the nucleons, the nucleons keep their identity intact except for the $\Delta S=1$ reaction where a nucleon is converted into a hyperon(in the case of $\bar\nu$ only). In the inelastic region the scattering leads to the excitation of various resonances. 
 The resonance excitation of the nucleon includes isospin 
$I=\frac{1}{2}$~resonances like $N^*(1440)$, $N^*(1520)$, $N^*(1535)$, etc., and isospin $I=\frac{3}{2}$~resonances like $\Delta(1232)$, $\Delta^*(1600)$, $\Delta^*(1700)$, etc., together with a non-resonant continuum. The decays of these resonances lead predominantly to single pion i.e. $\pi N$ state, and also to other final states like $\gamma N$, $\eta N$, $K Y$, $\pi \pi N$, $\rho N$, etc. 
 The shallow inelastic region(SIS) covers resonance excitation on the nucleon that, together with a non-resonant continuum, leads predominantly to the above mentioned final states. 
In Figure \ref{fig3}, we show the importance of the different kinematic regions relevant for the QE, IE and DIS scattering corresponding to the two neutrino energies viz. $E_\nu$ = 3 GeV (left panel) and E$_\nu$ = 10 GeV (right panel).
It can be observed from the figures that as one moves to the higher $\nu$ and $Q^2$ regions, the DIS becomes the most dominant
 process in the 
neutrino interactions where (anti)neutrino interacts with the 
quarks and gluonic degrees of freedom in the nucleons~\cite{SajjadAthar:2020nvy}. The DIS process, in this kinematic region, is described using perturbative QCD. However, presently there is no sharp kinematic boundary on $\nu$ and $Q^2$ for the onset of deep inelastic scattering in literature. Generally $Q^2 > 1GeV^2$ is chosen for the onset of DIS.
A kinematic
constraint of $W >$ 2 GeV is also applied to safely describe the DIS region. However, in the 
kinematic region of Q$^2 < 1 GeV^2$, nonperturbative QCD effects must be taken into serious
consideration. In this region, which is also known as the transition region, it is expected that 
the principle of
quark–hadron duality can be used to obtain the neutrino cross sections. However, there is not much work done either theoretically or experimentally to understand neutrino cross section using quark-hadron duality. This issue has been recently raised in the Snowmass~\cite{snowmass} and NUSTEC~\cite{NuSTEC:2019lqd} meetings.

\begin{figure}[h]
 \begin{center}
\includegraphics[width=5.4cm,height=8.5cm]{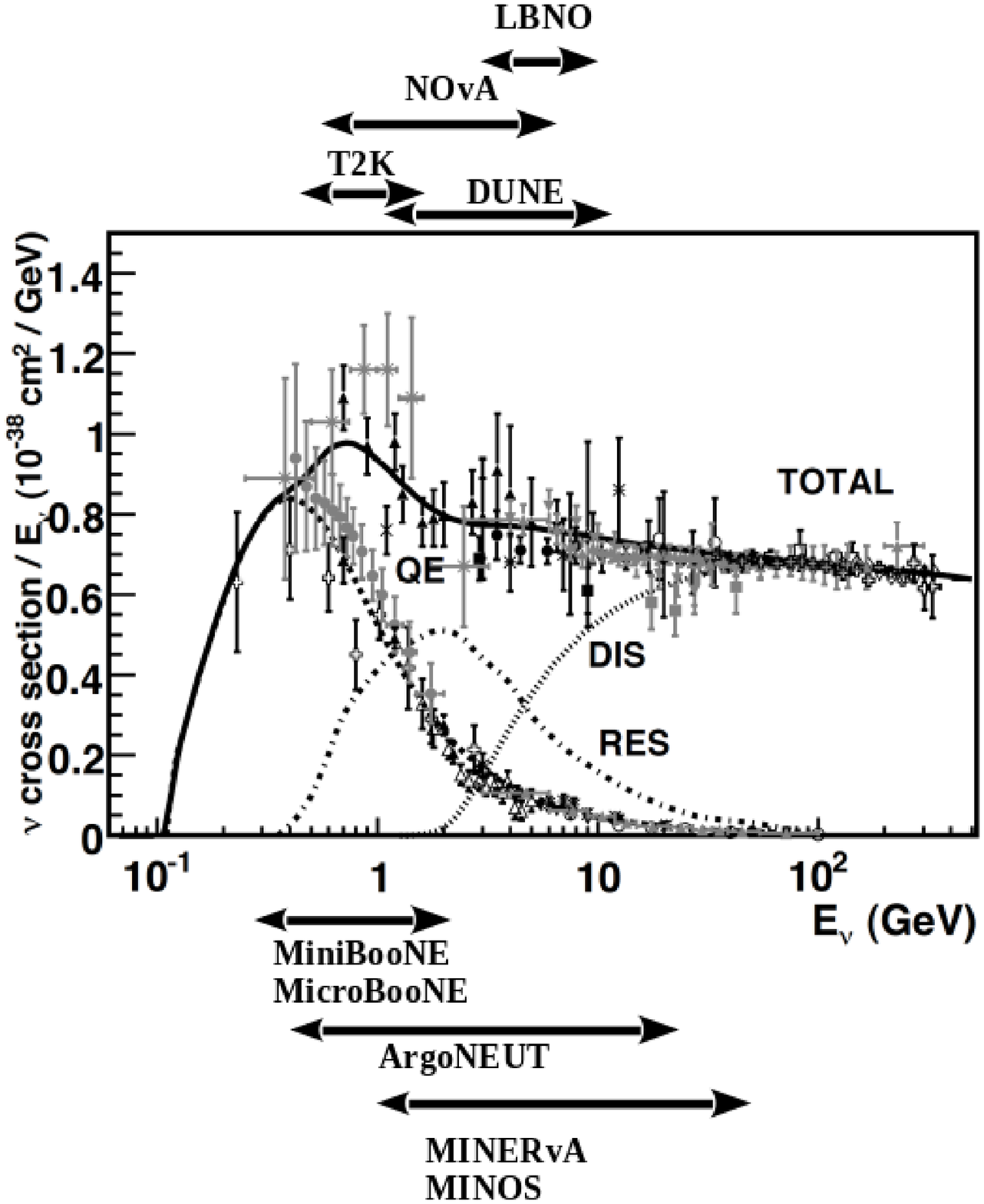}
\includegraphics[width=5.4cm,height=8.5cm]{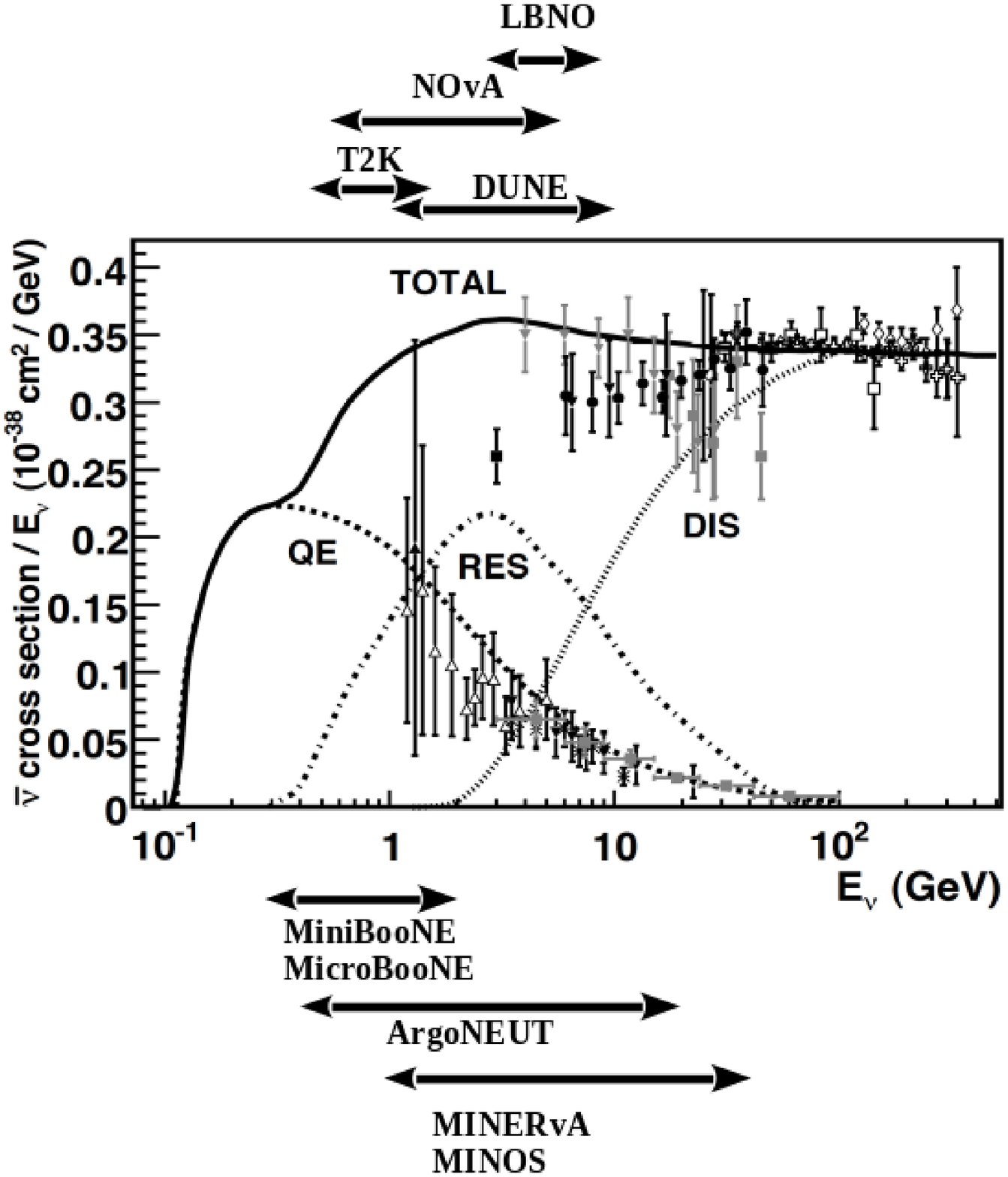}
\end{center}
\caption{Charged current induced total scattering cross section per nucleon per unit energy of the incoming particles vs. 
neutrino~(left panel) and antineutrino~(right panel) energy for all the three processes labelled on the curve along with the total scattering cross sections. Dashed line shows the contribution from the quasielastic~(QE) 
scattering while the dashed-dotted and dotted lines represent the contributions from the inelastic resonance~(RES) and deep 
inelastic scattering~(DIS), respectively. The sum of all the scattering cross sections~(TOTAL) is shown by the solid 
line~\cite{Lipari:1994pz}. We have also mentioned the energy region of various experiments.}
\label{zeller}
\end{figure}

 In Figure \ref{zeller}, we show the relative importance of the above processes(QE, IE and DIS) through the energy dependence 
 of their cross sections~\cite{Lipari:1994pz}.  
This figure depicts the total scattering cross section per nucleon per unit energy of the incoming particles vs. 
neutrino~(left panel) and antineutrino~(right panel) energy in the charged current induced process. The dashed line, 
dashed-dotted and dotted lines represent the contributions from the quasielastic scattering, inelastic resonance~(RES) and deep 
inelastic scattering~(DIS) processes, respectively. The sum of all the scattering cross sections~(TOTAL) is shown by the solid 
line~\cite{Lipari:1994pz}. The experimental results are the data from the older (ANL and BNL) as well as the experiments performed 
recently using (anti)neutrino beam. It may be realized that the experimental error bars are large and precise measurements are needed. In all the present generation neutrino experiments, nuclear targets like $^{12}$C, $^{16}$O, $^{40}$Ar, $^{56}$Fe, $^{208}$Pb, etc. are being used and the 
interactions take place with the nucleons that are bound inside the nucleus, where nuclear medium effects become important.

 These neutrino experiments are measuring (anti)neutrino events that are a convolution of
\begin{itemize}
 \item [(i)] energy-dependent neutrino flux and
 \item [(ii)] energy-dependent neutrino-nucleon cross section.
\end{itemize}
 Therefore, it is highly desirable to 
 understand the energy dependence of the neutrino-nucleon cross sections. In the context of the present neutrino oscillation experiments using the nuclear targets, it is of great importance to understand the energy dependence of nuclear medium effects.
 Especially in the precision era of neutrino oscillation experiments, 
to achieve an accuracy of a few percent~(2-3\%) in the systematics, a good understanding of neutrino-nucleon and neutrino-nucleus cross sections is required. Presently, due to the lack in the understanding of these cross sections, an uncertainty of 25--30\% in the systematics arise.   The study of neutrino interactions with matter is not only important for the understanding of neutrino physics but is also important to get a better insight of hadronic interactions in the weak sector, where there is an additional contribution of axial vector current besides the vector current.

In the case of QE process the general consideration of the nuclear medium effects(NME) include the Fermi motion, Pauli blocking and the multinucleon 
correlation effects. In the case of single pion production, one considers Fermi motion and Pauli blocking of the nucleon as well as the modification of the properties of the various excited resonances especially their masses and widths in the medium. However, these  modifications are well studied only in the case of $\Delta$ resonance. In addition, the pion produced in the decay of these resonances
undergo final state interaction with the residual nucleus, where charge 
exchange process(like $\pi^- p \rightarrow \pi^o n$) or modulation in its energy and momentum or pion absorption($\pi$ NN $\rightarrow$ NN) may take place. If the produced pion is absorbed it mimics a quasilike event. In the case of DIS, shadowing and antishadowing corrections become important in the region of low x, the Bjorken variable. In the intermediate region of x, the mesonic contributions  become important where the interaction of an intermediate vector boson(W,Z) takes place with the virtual mesons in the nucleus and  in  the region of high x, Fermi motion effects are important. 

The consideration of different NME is model dependent and there is no consensus on any particular nuclear model~\cite{NuSTEC:2017hzk,Katori:2016yel}.
In order to understand nuclear medium effects in (anti)neutrino-nucleus scattering, more data with better precision are needed.

This volume is devoted to the study of neutrino interactions from the nucleons and nuclei in the region of intermediate and high energies and comprises of seventeen articles discussing quasielastic, one pion production, other inelastic processes, and the 
deep inelastic scattering of (anti)neutrino from the nucleons and nuclei. All the contributing articles are arranged 
according to the following aspects of their contents:
\begin{itemize}
 \item [(i)] Experimental,
\item [(ii)] Theoretical, and
 \item [(iii)] Phenomenological.
\end{itemize}

To give a general historical understanding of neutrino experiments starting from Gargamelle to MINERvA, a bird's eye view has been lucidly illustrated by Morfin~\cite{Morfin}, 
where he summarizes various attempts which have been made for exploring the structure of the nucleon with neutrinos. In the next five articles, the current status and results of some important experiments being performed in the few GeV energy region are discussed like the efforts of 
MINERvA, NOvA, MicroBooNE, ArgoNeuT and 
neutrino interaction physics in neutrino telescope. The MINERvA@Fermilab took data using the (Anti)Neutrinos at the Main Injector (NuMI) beamline from 2009 to 2019 in the Low-Energy range and the Medium-Energy range that peak at 3 GeV
and 6 GeV, respectively, using several nuclear targets like carbon in scintillators, oxygen in water, iron and lead and the aim was to understand nuclear medium effects in the wide range of Bjorken $x$ and $Q^2$. Xianguo Lu et al.~\cite{MINERvA} on behalf of the MINERvA collaboration, have presented the latest 
results of the differential and total scattering cross sections for the inclusive, quasielastic, inelastic one pion, single kaon, etc. processes and highlights their salient observations. The NOvA@Fermilab has been collecting data in the NuMI
neutrino beam since 2014 and the expectation is that it will continue till 2026. Shanahan and Vahle~\cite{NOvA}, on behalf of the NOvA
 collaboration have presented experimental results in 3 flavor neutrino oscillation scenario as well as the results of the differential scattering cross sections for the inclusive channel and for the coherent pion production processes. The MicroBooNE and ArgoNeuT Liquid Argon 
Time Projection Chambers(LArTPC)@Fermilab have collected data in the NuMI and Booster Neutrino Beams, respectively.  
 The neutrino interaction measurements of these experiments are presented by Duffy et al.~\cite{MicroBooNE}, for the charged-current $\nu_\mu$ scattering in
the inclusive channel, $0\pi$ channel (in which no pions but some number of protons
may be produced), and  for the single pion production (including production of both charged
and neutral pions), as well as measurements of inclusive scattering cross sections for $\nu_e + \bar\nu_e$ interactions have been presented.
 Katori et al.~\cite{IceCube} have discussed the neutrino interaction physics in neutrino telescopes, where interactions are detected via Cherenkov radiation emitted by the charged secondaries, and specifically discussed in detail, the largest neutrino telescope in operation till date i.e. the IceCube Neutrino Observatory. 
 
 These articles are followed by ten articles dealing with various aspects of theoretical developments in the elastic, quasielastic, inelastic and the deep inelastic scattering of (anti)neutrinos from the nucleons and nuclei. Benhar~\cite{Benhar}
has elucidated the problems and uncertainties while evaluating the cross sections with nuclear medium effects and also deals with the theoretical understanding required to unravel the flux-averaged neutrino-nucleus cross  section by discussing in detail quasielastic scattering, one pion production and in very brief the deep inelastic scattering processes.
   Nuclear model dependence in quasielastic scattering has been discussed by Amaro et al.~\cite{Amaro} where they explicitly describe the neutrino-nucleus scattering cross section using superscaling (SuSA) approach by considering one- and two- body currents and showing first that
    the model explains well the electron scattering data and then applied it to understand weak interaction induced processes. Jackowicz and Nikolakopoulos~\cite{Jackowicz} have studied nuclear medium effects in quasielastic neutrino-nucleus scattering by using nuclear mean field and random phase approximation, and highlighted the differences between neutrino- and
antineutrino-induced reactions. Martini et al.~\cite{Martini} have described the neutrino-nucleus scattering cross section for CCQE process using response functions as well as spectral functions, and highlighted 
 the model dependence of multinucleon correlation effects in the different models. Alvarez-Ruso et al.~\cite{Alvarez-Ruso} have discussed neutrino interactions with matter with the particular emphasis on the 
 MiniBooNE anomaly. Fatima et al.~\cite{Fatima} have pointed out the importance of $\bar{\nu}_{\mu}$ induced quasielastic production of hyperons leading to pions(the reaction which is forbidden for the neutrino induced processes due to $\Delta S=\Delta Q$ rule). The effects of the second class currents in the axial vector sector with and without T-invariance as well as the effect of SU(3) symmetry breaking have also been discussed. Paschos~\cite{Paschos} has discussed a model for the flavor changing neutral current of leptons.
 
 Neutrino-nucleon reactions in resonance region have been studied by Sato~\cite{Sato} using dynamical coupled channel(DCC) model by restoring full unitarity. The cross sections of charged current neutrino reaction are examined to analyze the
mechanism of the neutrino induced meson production reaction, and possible way to
test the model of the axial vector current contribution. In the case of deep inelastic scattering, the evolution of the electroweak structure functions of nucleons has been studied by Reno~\cite{Reno}
 in the context of muon and tau neutrino and antineutrino scattering. Ansari et al.~\cite{Ansari} in their review article have discussed the effect of the nonperturbative corrections such as the 
 target mass correction and higher twist effects, perturbative evolution of the parton densities, nuclear medium modifications of the nucleon structure functions, nuclear isoscalarity corrections on the weak nuclear structure functions in the (anti)neutrino-nucleus scattering in the DIS region. The numerical results for the structure functions and the cross sections are compared with some of the available experimental data including the recent results from MINERvA. The predictions are made in argon nuclear target which is planned to be used as a target material in DUNE at the Fermilab.
 
Neutrino event Monte Carlo(MC) generators play an important role in the design, optimization, and execution of neutrino oscillation 
experiments and the two 
most widely used  neutrino event generators in the present experimental physics community are GENIE and NEUT which predict the neutrino event rates by using the various inputs including the (anti)neutrino-nucleus scattering cross sections and the final state interactions 
of the produced hadrons in the nucleus. In this volume main features of these two generators are separately elaborated. GENIE as explained by Alvarez-Ruso et al.~\cite{GENIE}, with its gradual evolution and adaptability is assumed to become a standard tool, forming an
indispensable part of many experiments and has been widely tested against neutrino cross-section data. Important features of the NEUT MC generator have been illustrated by Hayato et al.~\cite{NEUT} It can be used to
simulate interactions for neutrinos between 100 MeV and a few TeV of energy, and is also 
capable of simulating hadron interactions within a nucleus. 

 We are thankful to all the authors who have contributed to this volume. Really their efforts are commendable and we hope that this volume will be helpful to the students, young as well as senior research workers in the field of neutrino physics and would stimulate some new ideas and investigation. Special thanks are due to B. Ananthanarayan, member of the editorial board of EPJ-ST who invited us to bring this topical volume.  The help and cooperation of the editorial team of EPJ-ST is duly acknowledged.

\end{document}